# On understanding the figures of merit for detection and measurement of x-ray polarization

Martin C. Weisskopf[1 a], Ronald F. Elsner[a], Stephen L. O'Dell[a]

[a]NASA Marshall Space Flight Center, Space Science Office, Huntsville, AL 35812 USA

## ABSTRACT

The prospects for accomplishing X-ray polarization measurements appear to have grown in recent years after a more than 35-year hiatus. Unfortunately, this long hiatus has brought with it some confusion over the statistical uncertainties associated with polarization measurements of astronomical sources. The heart of this confusion stems from a misunderstanding (or potential misunderstanding) of a standard figure of merit—the minimum detectable polarization (MDP)—that one of us introduced many years ago. We review the relevant statistics, and quantify the differences between the MDP and the uncertainty of an actual polarization measurement. We discuss the implications for future missions.

## 1. INTRODUCTION

Only a few experiments have conducted X-ray polarimetry of cosmic sources. In several rocket observations, we measured X-ray polarization from the Crab Nebula[1]. Using the X-ray polarimeter on the Orbiting Solar Observatory (OSO-8), we confirmed this result with a 19-sigma detection ($P$ = 19.2% ± 1.0%)[2], thus proving the synchrotron origin of the X-ray emission from this object. Data from OSO-8 also provided upper limits to polarization of a handful of bright targets. The key challenge for broad-band X-ray polarimetry has always been to obtain an instrument that is highly responsive to polarized radiation yet has low net background. In the past decade the development of a class of polarimeters based on the photoelectric effect[3,4] and placed at the focus of an X-ray telescope has produced a renewed interest.

## 2. STATISTICS

All past and currently planned X-ray polarimeters for astronomical applications share the feature that ultimately the data may be described as binned as a function of the azimuth (φ) around the direction of the line of sight to the source. Polarization manifests itself as modulation at (2φ). Often polarimeters are rotated around the line of sight to address potential systematic effects; thus one commonly uses the language that polarization will manifest itself as a modulation of the detected signal at twice the rotation frequency. Here both amplitude and phase information are important as the former is related to the degree of polarization and the latter to the position angle.

In the presence of noise, detecting this type of signal is most often discussed in classical post-war textbooks in the context of modulated radio waves in the presence of noise[5]. In the special case where the noise is described by Poisson distribution and $N$ is the total number of counts, the probability $p(a,\varphi)$ of measuring a particular modulation amplitude $a$ and a phase $\varphi$, given that the true amplitude and phase are $a_0$ and $\varphi_0$, is as follows:

$$p(a,\varphi) = \frac{N\,a}{4\pi}\exp\left[-\frac{N}{4}[a^2 + a_0^2 - 2aa_0\cos(\varphi-\varphi_0)]\right]. \tag{1}$$

[1] martin.c.weisskopf@nasa.gov; phone +1(256)961-7798;
NASA/MSFC/VP62, 320 Sparkman Drive, Huntsville, AL 35805-2718 USA

It should be readily apparent, since the modulation is positive definite, that there is *always* a probability of measuring an amplitude of modulation, even if the underlying source is unpolarized ($a_0$=0). Equation (1) may be integrated analytically and inverted to determine the amplitude which has a given probability of being exceeded by chance for an unpolarized ($a_0$=0) source. For example, the amplitude, $a_{1\%}$, that has only a 1% probability of being exceeded by chance is

$$a_{1\%} = \frac{4.29}{\sqrt{N}}. \tag{2}$$

One is not directly interested in the amplitude of modulation but the degree of polarization that corresponds to. There are two steps in relating these. The first step is recognizing that there may be a contribution from a steady background to the detected signal and therefore to express the modulation as a fraction of the detected signal only:

$$a_S = a_{1\%} \times (R_S + R_B) / R_S . \tag{3}$$

Here $R_S$ and $R_B$ are the signal and background counting rates. The next step introduces the modulation factor, $M$, the amplitude of modulation for a 100% polarized input *in the absence of background*. The modulation factor may be a function of energy. In the typical photoelectric polarimeter, $M$ will be small at the low energies where tracks are short and the angular distribution appears almost spherical, but will rises very quickly above these energies to values possibly approaching 0.5. We define the minimum detectable polarization (MDP) as ($a_s/M$):

$$MDP = \frac{a_S}{M} = \frac{4.29}{M\,R_S}\left[\frac{R_S + R_B}{T}\right]^{1/2}. \tag{4}$$

By this definition, the "minimum detectable polarization" is the degree of polarization corresponding to the amplitude of modulation that has only a 1% probability of being detected by chance. The MDP, however, is *not* the uncertainty of a polarization measurement. This uncertainty, at a certain level of confidence, is found by integrating Equation (1):

$$C = \iint a\, da\, d\varphi\; p(a,\varphi). \tag{5}$$

Here $C$ is the desired confidence level ($0 \le C \le 1$) and the integral is performed over an appropriate sub-region of the parameters space ($a, \varphi$). To invert Equation (5), we begin by defining the Stokes amplitudes:

$$\begin{aligned} a_{0,x} &= a_0 \cos \varphi_0, \\ a_{0,y} &= a_0 \sin \varphi_0, \\ a_x &= a \cos \varphi = a_{0,x} + \Delta a \cos \psi, \\ a_y &= a \sin \varphi = a_{0,y} + \Delta a \sin \psi \end{aligned} \tag{6}$$

Here ψ is a parametric variable that we shall vary between $0$ and $2\pi$ and

$$\Delta a^2 = a^2 + a_0^2 - 2\,a\,a_0 \cos(\varphi - \varphi_0). \tag{7}$$

The amplitude $a$ and phase angle $\varphi$ on a given confidence contour may then be expressed terms of $a_0$, $\varphi_0$, $\psi$, and $\Delta a = \Delta a_C$ as

$$\begin{aligned} a &= (a_0^2 + \Delta a_C^2 + 2a_0\,\Delta a_C \cos(\psi - \varphi_0))^{1/2}. \\ \varphi &= \arctan\left(\frac{a_0 \sin\varphi_0 + \Delta a_C \sin\psi}{a_0 \cos\varphi_0 + \Delta a_C \cos\psi}\right). \end{aligned} \tag{8}$$

Using the relationship $a\, da\, d\varphi = d\Delta a_x\, d\Delta a_y = \Delta a\, d\Delta a\, d\psi$, Equation 5 may be expressed as

$$C = \int_0^{\Delta a_C} \int_0^{2\pi} \Delta a\, d\Delta a\, d\psi \frac{N}{4} e^{-N\,\Delta a^2/4}\ . \tag{9}$$

The integrals in Equation 9 may be evaluated and solved for $\Delta a_C$:

$$\Delta a_C = \sqrt{-\frac{4}{N}\ln(1-C)}\ . \tag{10}$$

Equations 8 and 10 then may be used to draw the appropriate confidence contours.

## 3. HOW MANY COUNTS ARE NECESSARY?

We are now in a position to examine the answer to two questions: (1) How many counts are required to achieve a particular minimum detectable polarization? (2) How many counts are required to *measure* this same level of polarization to a given accuracy? To simplify the discussion we set the background level to zero. (Allowing for the background we leave as an exercise for the reader.) With no background, Equation 4 simplifies and may be solved for the number of counts, $N_1$ necessary to achieve a particular *MDP*:

$$N_1 = \left(\frac{4.29}{M \times MDP}\right)^2 . \tag{11}$$

For example, $N_1 = 1{,}565{,}369$ for a modulation factor of *1/3* and an MDP of *0.01 (1%)*. Next we want to answer the question how many counts are required to measure the amplitude of modulation corresponding to this MDP to say 3, 4, 5, …, β-sigma accuracy? We have determined the answer to this question in two ways. In the first, we made use of the system of equations (8) and (10) to determine the extremes in the amplitude of modulation on the 67% confidence contours. We then found the number of counts, $N_2$, such that

$$\frac{a_0}{(a_{max} - a_{min})/2} = \beta\ . \tag{12}$$

Where β =1, 2, 3, 4, … is an approximate proxy for the number of sigma of the measurement. We find that the ratio, $N_2/N_1$, scales as $\beta^2$ and with a coefficient that for all practical purposes is ¼ as shown in Figure 1. As a cross-check we performed the following derivation:

$$a_{MDP}(N_1) = 4.29/\sqrt{N_1}\ ;$$

$$\Delta a(C, N_2) = 2\sqrt{-\frac{\ln(1-C)}{N_2}}\ . \tag{13}$$

Finally, let $a_{max,C} - a_{min,C} = \eta\, \Delta a(c,N_2)$; then

$$\left(\frac{N_2}{N_1}\right) = -\left(\frac{\beta\,\eta}{4.29}\right)^2 \ln(1-C) \ , \tag{14}$$

which reproduces Figure 1 for η=4.29/2 and C=0.67.

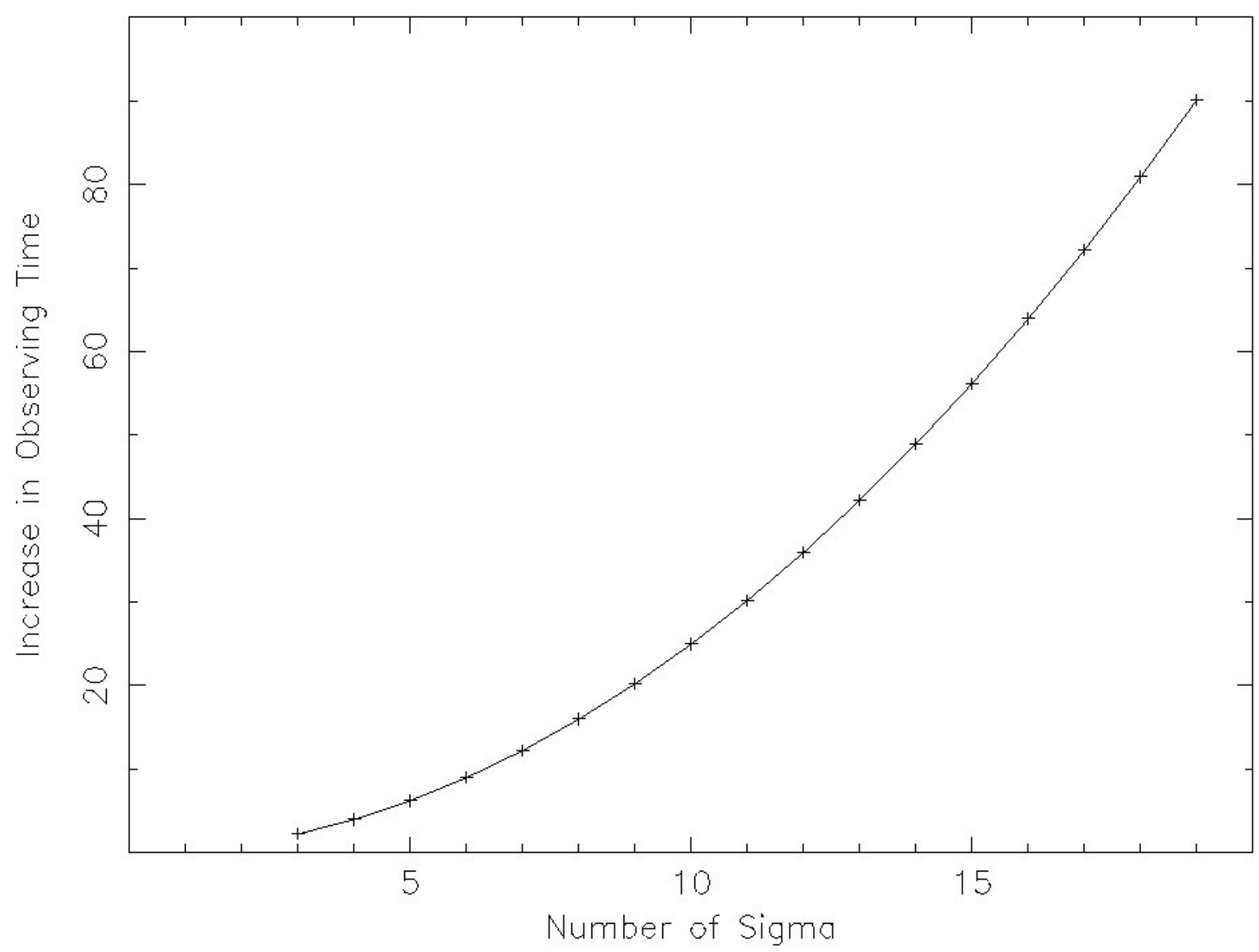

Figure 1. The increase in required observing time ($N_2/N_1$) versus the number of sigma ($\beta$) in order accurately to measure the polarization.

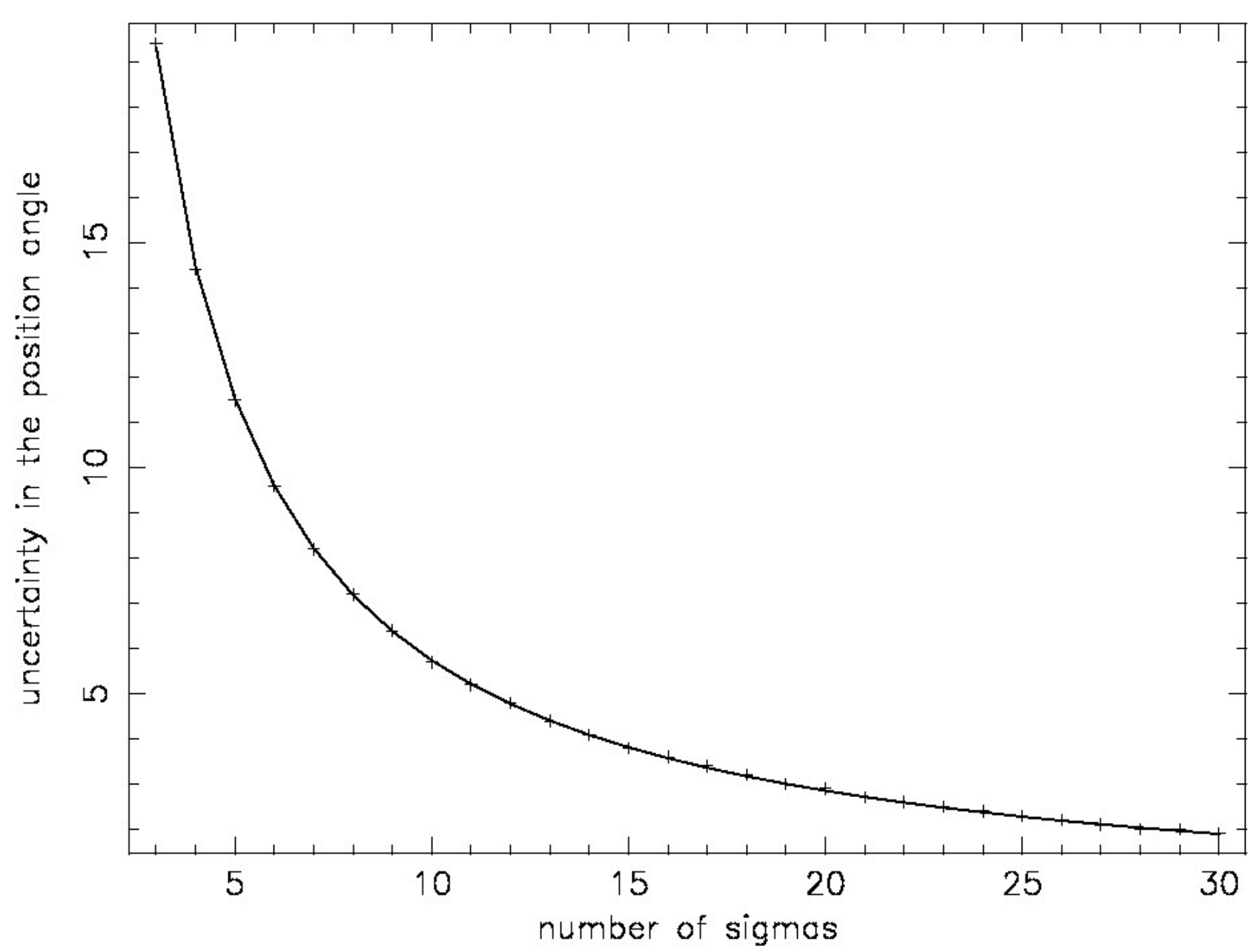

Figure 2. The 1-σ uncertainty in the position angle versus the number of sigma ($\beta$) in figure 1.

For completeness, we show in Figure 2 an estimate of the uncertainty in the position angle based on ½ the extremes on the 1-sigma contours. This estimator scales as $1/\beta$. As with the degree of polarization, the conclusion that a precise measurement of the position angle requires far more counts than those necessary to determine the *MDP* is inescapable.

## 4. SUMMARY AND CONCLUSION

Here we have reviewed the statistical description appropriate for an X-ray polarimeter that attempts to measure polarization by searching for a variation of the detected signal at twice the rotation frequency. Our analysis applies to the case where the signal is dominated by Poisson statistics and our examples ignore the background. We have explicitly shown that the commonly used figure of merit for such a polarimeter—i.e., minimum detectable polarization (MDP)—is *not the uncertainty in the polarization measurement*. In particular, we compared the MDP at 99% confidence with a nominal measurement at the β-sigma level: We have shown that the number of counts required to measure the polarization is larger than that required to establish the 99% MDP, by a factor of $\beta^2/4$! Thus, a proper *measurement*, even at the 3-sigma level, requires an integration time at least 2.25 times longer than that simply to establish that one might have detected polarization at the level set by the $MDP_{99\%}$. A more reasonable criterion may be that the measurement should be accurate to 4- or 5-sigma if one wants to discriminate amongst different theoretical models for the source of the polarization: This requires an increase in observing time by a factor of 4 or 6.25, respectively. The implication is that observing programs where the integration time is set by the MDP may significantly underestimate the time required for *measurement* of polarization. This is especially true in those cases where there are grounds (prior measurements, physics, and observations at other wavelengths) to expect a certain degree of polarization to be present. For example, the degree of accuracy of polarization measurements as a function of pulse phase for the Crab pulsar may be estimated from optical measurements.